\documentclass[useAMS,usenatbib]{mn2e}
\usepackage{array,multirow,epsfig,graphicx,amssymb,amsmath,lscape}
\usepackage{txfonts}
\usepackage{aas_macros}

\newcommand{\HH}{\mathrm{H}}
\newcommand{\Hmol}{\mathrm{H}_2}
\newcommand{\ZZ}{\mathrm{z}}
\newcommand{\HE}{\mathrm{He}}

\newcommand{\sound}{\mathrm{s}}
\newcommand{\Ross}{\mathrm{R}}
\newcommand{\Planck}{\mathrm{P}}
\newcommand{\gas}{\mbox{g}}
\newcommand{\AU}{\mbox{ AU}}
\newcommand{\g}{\mbox{ g}}
\newcommand{\s}{\mbox{ s}}
\newcommand{\K}{\mbox{ K}}
\newcommand{\cm}{\mbox{ cm}}
\newcommand{\rad}{\mbox{r}}
\newcommand{\CC}{\mathrm{c,H}}
\newcommand{\VSG}{\mathrm{KSG}}
\newcommand{\NSG}{\mathrm{NSG}}
\newcommand{\eff}{\mbox{\small{eff}}}
\newcommand{\yr}{\mathrm{yr}}
\newcommand{\Msun}{\mathrm{M}_\odot}
\newcommand{\revs}[1]{{#1}}
\defcitealias{2004ApJ...603..383T}{TM04}
\title{Stationary population {\sc iii} accretion discs}

\author[Mayer \& Duschl]{Michael Mayer$^1$ and Wolfgang J.\ Duschl$^2$\\
Institut f\"ur Theoretische Astrophysik, Tiergartenstr. 15,
69121 Heidelberg, Germany\\
$^1$ E-Mail: mm@ita.uni-heidelberg.de $^2$ Email:
wjd@ita.uni-heidelberg.de}

\begin{document}
\date{\today }

\pagerange{\pageref{firstpage}--\pageref{lastpage}} \pubyear{2004}

\maketitle

\label{firstpage}

\begin{abstract}
We present stationary models of protostellar population {\sc iii}
(Pop {\sc iii}, for short) accretion discs, compare them to Pop {\sc i}
discs, and investigate the influence of the different chemical
compositions on the occurence of gravitational, thermal and
thermal-viscous instabilities in the discs. In particular in the
cooler regions, we find major differences between Pop {\sc
iii} and Pop {\sc i} discs, both in the structure and stability
behaviour. This is mainly due to the absence of most molecules and
dust in Pop {\sc iii}, which are very efficient absorbers in Pop
{\sc i} discs.
\end{abstract}


\section{Introduction}

The important role of angular momentum during stellar and
planetary formation is best evidenced by the remarkable
co-planarity of the planetary orbits in our solar system.
Observations of protostellar and protoplanetary accretion discs
\citep{Hartmann98} point towards this being generally the case.
Even in the lack of direct evidence it is reasonable to assume
that---as far as the role of angular momentum is concerned---the
situation was not substantially different for the very first
generation of stars, the population {\sc iii} (Pop {\sc iii}, for short).

While the general role of angular momentum may have been the same,
and certainly will have led to the formation of population {\sc
iii} accretion discs, the structure and the stability of these
discs may have differed considerably from that of contemporary
protostellar accretion discs. This difference in structure and
evolution is not surprising given the difference in the chemical
composition of accretion discs now (Pop {\sc i}) and then (Pop
{\sc iii}), in particular the lack of efficient cooling agents in
the primordial matter discs.

Models of accretion in early universe so far assumed spherical
accretion \citep{1986ApJ...302..590S}. Only recently people
started to work on accretion through a disc \citep[][herafter
\citetalias{2004ApJ...603..383T}]{2004ApJ...603..383T}. In
particular, the importance of the opacity coefficient for the
stability of accretion discs has been known for quite a while
\citep[e.g.][]{Duschl93, 2001A&A...367.1087H}. In this
contribution, we wish to investigate the structure and stability
of population {\sc iii} accretion discs. For this purpose, we
present stationary, vertically integrated, axially symmetric discs
models.

In Sect.\ \ref{Sect:model} we shall summarize the properties of
the accretion disc model we used and present the set of equations
solved. The subsequent Sect.\ \ref{Sect:numerics} is dedicated to
a short summary of the numerical method used to solve these set of
equations. We then (Section \ref{Sect:opacity}) discuss the
opacity coefficient for a Pop {\sc iii} chemical composition.
Before we proceed to the presentation of the accretion disc models
(Sect.\ \ref{Sect:results}), we shortly review criteria for
thermal, thermal-viscous and gravitational instabilities in
Section\ \ref{Sect:stability} and give formulae for the timescales
present in discs (Sect.\ \ref{Sect:timescale}) developing handy
formulae for relevant limiting cases. We end in Sect.\
\ref{Sect:discussion} with discussions of these results and put
them into perspective.

\section{The accretion disc model} \label{Sect:model}

We use a standard one zone accretion disc model
\citep{1981ARA+A..19..137P} with slight but important
modifications compared to the results of
\cite{2003PYunO..95..151M}: Our discs are allowed to be vertically
self-gravitating and can become optically thin. Vertical
selfgravity means that in a direction perpendicular to the
rotational plane of the disc the gravitational acceleration is no
longer restricted to be due to the projected contribution from the
central object only, but now can also be due to the local mass
distribution, or both at the same distance from the central
object. The gravitational pull in radial direction is still only
due to the central force as we exclude solutions where the disc
mass becomes larger than the central mass, i.e. the rotational law
is still Keplerian \citep[KSG discs, Keplerian self-gravitating,
see ][]{2000A&A...357.1123D}.

We have

\begin{itemize}
\item the continuity equation
\[
\dot M = 2\pi r \Sigma u_r
\]
\item the angular momentum equation ($\nu$ viscosity; $f(r)$
accounts for the inner boundary condition, see
eq.~\ref{eq:innerboundary})
\begin{equation}
\nu \Sigma = \frac{\dot M}{3\pi}f(r) \label{eq:angmom}
\end{equation}
\item the energy equation 
\begin{equation}
Q^+=\frac{9}{8}\nu \Sigma\frac{GM}{r^3}=\frac{4\sigma
}{3\tau_{\eff}}T^4=Q^-\label{eq:energy}
\end{equation}
It has been shown \citepalias{2004ApJ...603..383T} that the
ionisation energy may not be neglected although the thermal energy
is negligible. We do not account for this effect in our models
(see discussion in Appendix \ref{Sect:tmc}).

\item hydrostatic equilibrium in vertical direction
\begin{equation}
\frac{1}{\rho}\frac{P}{H}=g_{\CC}+g_{\VSG}=2G\left(\frac{M}{4\rho
r^3}+\pi\right)\Sigma\label{eq:pressure}
\end{equation}
The pressure due to matter and radiation is sustained by the
vertical component of the gravitational force $g_{\CC}$ of the
central body and the vertical selfgravity of the disc
($g_{\VSG}=2\pi G \Sigma$, \cite{1978AcA....28...91P})
\end{itemize}

The optically thick $\rightarrow$ optically thin transition is
achieved by modifying the equation of state and the optical depth
\citep{1996ApJ...456..119A}
\begin{equation}
P=P_{\gas}+P_{\rad}=\rho\frac{kT}{\mu
m_{\mbox{p}}}+\frac{4\sigma}{3c\tau_{\eff}}T^4\left(\tau_{\Ross}+\frac{4}{3}\right)
\label{eq:pressure2}
\end{equation}

The effective optical depth is given by
\begin{equation}
\tau_{\eff}=\tau_{\Ross}+\frac{4}{3}+\frac{2}{3\tau_{\Planck}}\label{eq:tau}
\end{equation}
where we define the Rosseland and Planck optical depth by
\begin{equation} \label{eq:rossplanck}
\tau_{\Ross/\Planck}=\frac{1}{2}\Sigma \kappa_{\Ross/\Planck}(\rho,T)
\end{equation}
The radiation pressure reaches the black-body value
$P=\frac{4\sigma}{3c}T^4$ for $\tau_{\eff}=\tau_{\Ross}\rightarrow
\infty$ and is negligible in the optically thin regime
($\tau_{\Ross}<1$, $\tau_{\eff}=\frac{2}{3\tau_\Planck}\rightarrow
\infty$, $\frac{\tau_{\Ross}+\frac{4}{3}}{\tau_{\eff}}\rightarrow
0$). In a optically thin medium photons are not able to contribute
to the pressure in the gas as the mean-free path becomes long
compared to the physical size of the medium they are passing
through. The formal optical depth in eq. (\ref{eq:tau}) which
enters into the energy eq. (\ref{eq:energy}) equals the usual
diffusion approximation for $\tau_\Ross \gg 1$ and moves into the
optical thin regime for $\tau_\Ross \ll 1$. In this respect we
account for the transition from the optically thick diffusion
approximation ($Q^-=\frac{4\sigma}{3\tau_\Ross}T^4$) to the
optically thin emission ($Q^-=2\sigma T^4\tau_\Planck$.

The inner boundary condition ($\Sigma_{\rm in}=0$) can be expressed by
\begin{equation}
  \label{eq:innerboundary}
  f(r)=1-\sqrt{\frac{r_*}{r}}
\end{equation}
Once we are sufficiently away from $r_*$, $f(r)=1$. For our models
we set $f(r)=1$. (see Sect. \ref{Sect:selfsimilar})

We furthermore define 2 parameters
\[
\beta_P=\frac{P_{\gas}}{P_{\gas}+P_{\rad}}\qquad
\eta_\VSG=\frac{g_\CC}{g_{\CC}+g_\VSG}=\frac{M}{M+4\pi\rho
r^3}
\]
$\beta_P$ measures the contribution of the gas pressure to the
total pressure ($\beta_P=1$: pure gas pressure, $\beta_P=0$: pure
radiation pressure), and $\eta_\VSG$ the vertical
selfgravity ($\eta_\VSG=1$: no vertical selfgravity,
$\eta_\VSG=0$: fully vertically selfgravitating).

\begin{table}
\centering
\begin{tabular}{l|l}
\hline
$\dot M$ & accretion rate\\
$\Sigma$ & Surface density $\Sigma=2\rho H$\\
$\rho$ & density \\
$T$ & temperature \\
$H$ & disc half thickness \\
$\nu$ & viscosity prescription \\
$u_r$ & Radial drift velocity \\
$\Omega$ & Rotation frequency $\Omega_{\NSG/\VSG}=\sqrt{\frac{GM}{r^3}}$  \\
$P$ & pressure \\
$c_\sound$ & sound speed\\
$\tau_{\eff}$ & effective optical depth \\
$M$ & central mass \\
$r$ & radial distance \\
$\kappa_{\Ross/\Planck}$ & Rosseland and Planck opacity\\
$\mu$ & mean molecular weight\\
\hline
\end{tabular}
\caption{Nomenclature of the variables used. If not stated
otherwise, the symbols refer to the disc's
mid-plane}\label{tbl:symbols}
\end{table}

For the viscosity prescription we use $\alpha$-viscosity
\citep{1973A&A....24..337S}
\begin{equation}
\nu=\alpha \frac{P}{\rho\Omega} \label{eq:alphaP}
\end{equation}
This corresponds to a stress tensor element $t_{r\phi}=-\alpha_P
P$ with $\alpha_P=\frac{3}{2}\alpha$ which translates into the
more commonly used $\nu=\alpha c_{\sound} H$ for the
non-selfgravitationg case. We use this more general viscosity
prescription to avoid the difficulties of isothermality and
rapidly increasing density in the vertically and fully
selfgravitating parts of accretion discs when using $\nu=\alpha
c_{\sound} H$ in all cases \citep{2000A&A...357.1123D}\footnote{We
note that in \cite{2000A&A...357.1123D}'s notation the viscosity
parametrization (\ref{eq:alphaP}) corresponds to what they call
the dissipation-limited case, i.e., where the turbulent velocity
is limited by the sound velocity.}.

\section{Numerical treatment} \label{Sect:numerics}

As we do not treat radial selfgravity and therefore have no
implicit radial coupling, we are able to reduce the equations
above into two equations depending only on density and
temperature.
\begin{eqnarray}
G\left(\frac{M}{4\rho r^3}+\pi \right)\left(\frac{\dot M}{3\pi \nu}\right)^2
&=& \rho\frac{kT}{\mu m_{\mbox{p}}}+\frac{4\sigma
T^4}{3c\tau_{\eff}}\left(\frac{\kappa_{\Ross}\dot M}{6\pi
\nu}+\frac{4}{3}\right) \label{eq:eq01}\\
\frac{3}{8\pi}\frac{GM\dot M}{r^3}&=&\frac{4\sigma}{3\tau_{\eff}}T^4\label{eq:eq02}
\end{eqnarray}

We solve these equations using a 2D nested intervals method for
$(\rho,T)$ in the range of $-16 < \log \rho < -2$ and $1.8 < \log
T < 4.6$. We refine the initial rectangle 10 times so that we end
up with an accuracy of $\Delta\left(\log \rho\right)\approx 0.02$
and $\Delta\left(\log T\right)\approx 0.003$.

In our system of equations we have left the effective optical
depth $\tau_{\eff}$ (given by eq. \ref{eq:tau}) and the viscosity
$\nu$. For the $\alpha$-viscosity prescriptions it is impossible
to explicitly calculate $\nu\left(\rho,T\right)$ as the radiation
pressure depends on the surface density which itself depends on
the viscosity. This implies a polynomial equation for $\nu$ of
order 3. The details of this calculation are given in Appendix
\ref{sect:viscositydet}. The optical depth still containing the
surface density can be fully expressed in terms of density and
temperature by using the angular momentum equation
(\ref{eq:angmom}).

\section{Opacity} \label{Sect:opacity}

Primordial matter as an unprocessed relict from the big bang
strongly differs from today's chemical composition as then the
heavier elements which have to be produced first in stars were not
present. Due to the lack of these metals, no dust could form. This
lack of dust affects the opacities at low temperature, as dust is
a very strong absorber.

In the absence of good absorbers unusual absorption effects become
important: in the high-density and low-temperature regime
collision-induced absorption (CIA) of $\Hmol/\Hmol$, $\Hmol/\HE$,
$\Hmol/\HH$ and $\HH/\HE$-pairs dominates. In the low-temperature
and low-density regime we have Rayleigh scattering in the
Rosseland opacity while CIA continues to be the dominating effect
in the Planck mean.

As there was no consistent and complete set of Pop {\sc iii}
opacities available in the literature (except the ones of Lenzuni
et al. from 1000-7000 K and OPAL and OP ones above 10 000 K), we
computed them on our own (Mayer, Diploma thesis; Mayer \& Duschl,
in prep.). The results are Rosseland and Planck mean opacities for
$-16 < \log \rho < -2$ and $1.8 < \log T < 4.6$ assuming chemical
equilibrium and only including continuum absorption (no lines).
Our opacities match those of Lenzuni better than 10 \% above 3000
K. Below 3000 K deviations are due to the implementation of CIA
\cite[e.g.][]{1989ApJ...336..495B}\footnote{For an up-to-date
reference of available CIA data see {\tt
http://www.astro.ku.dk/}\~{\tt aborysow/programs/}}. The
consistency with the OPAL and OP opacities is of the same level of
accuracy, although we see larger differences in the high-density
limit due to neglected pressure ionization.

During the last years it became widely accepted that it is sufficient
to take hydrogen and helium species into account when dealing with
Pop {\sc iii} matter. However there is a small number fraction of
deuterium and lithium present, the latter of which must not be
neglected.

Lithium belongs to the group of Alkali metals and can therefore
easily be ionized. We included a simple deuterium and lithium
chemistry. The deuterium chemistry does not affect the opacity at
all, but the lithium changes the opacity up to two
orders of magnitude due to its ability to provide free electrons
at comparatively low temperatures. We do not include absorption effects of
lithium and deuterium explicitly, lithium is only needed to alter
the chemical equilibrium and therefore the absorption properties.
The largest positive differences relative to the $\mathrm{Z}=0$ calculation
occur at temperatures around a few 1000 K and at high and low
densities while there is a negative difference at intermediate
densities due to the destruction of $\HH_3^+$.

Our opacity calculations show a difference between true Pop {\sc
iii} and zero metallicity matter in this temperature range.
Currently we only included continuum absorption of hydrogen and
helium species with the exception of $\HH_3^+$ bound-bound
transitions. Further details, a discussion of the opacity
calculation and the inclusion of line absorption can be found in
Mayer \& Duschl (2004, in prep.). The models presented here were
calculated with a Lithium fraction of $f_{\mathrm{Li}}=5\cdot
10^{-10}$ and a Deuterium fraction of $f_{\mathrm{D}}=4\cdot
10^{-5}$.

To compare with Pop {\sc i} opacity we use the most recent
compilation of low-temperature dust and gas opacities by
\cite{2003A+A...410..611S}.

\begin{table}
\begin{center}
\begin{tabular}{|l|l|r|}
\hline \multirow{2}{1.3cm}{$P_{\gas}\gg P_{\rad}$} &
$\tau_{\Ross}\gg 1$ &
$A_{\Ross}-(1+\eta_{\VSG})B_{\Ross}+3(1+\eta_{\VSG})<0$
\\\cline{2-3}
 & $\tau_{\Planck}\ll 1$ & $-A_{\Planck}+(1+\eta_{\VSG})B_{\Planck}+3(1+\eta_{\VSG})<0$ \\\cline{1-3}
\hline \multirow{2}{1.3cm}{$P_{\rad}\gg P_{\gas}$} &
$\tau_{\Ross}\gg 1$ & $4A_{\Ross}-\eta_{\VSG} B_{\Ross}-4<0$
\\\cline{2-3}
 & $\tau_{\Planck}\ll 1$ & $-4A_{\Planck}+\eta_{\VSG} B_{\Planck}-4<0$ \\\cline{1-3}
\hline
\end{tabular}
\end{center}
\caption{Conditions for thermal instability} \label{tbl:thermal}
\end{table}

\section{Stability} \label{Sect:stability}

\subsection{Gravitational instability} \label{sect:toomre}
The gravitational instability is usually characterized by the
Toomre parameter \citep{1963ApJ...138..385T}:
\begin{equation}
\label{eq:toomre}
Q=\frac{\Omega c_\sound}{\pi G \Sigma}
\end{equation}
where we make use of $\kappa^2=4\Omega^2+2R\Omega
\frac{d\Omega}{dR}=\Omega^2$ being the epicyclic freqeuency for
Keplerian rotation. If $Q\ll 1$, then the disc might be prone to
axisymmetric instabilities leading to fragmentation. The true
causes for low-$Q$ discs are still under debate \citep[e.g. see
discussion in][]{2003MNRAS.339..937G}.

\subsection{Thermal instability} \label{Sect:thermalinstability}

Suppose we got an equilibrium solution from eqns. \ref{eq:eq01}
and \ref{eq:eq02}. Then the disc is thermally unstable if an
increase (decrease) of the heating rate is answered by a slower
increase (decrease) of the cooling rate. In mathematical terms
this criterion is given by
\begin{equation}
  \label{eq:stabthermal}
  \left.\left(\frac{\partial \log Q^+}{\partial \log T}\right)\right|_P-\left.\left(\frac{\partial \log Q^-}{\partial \log T}\right)\right|_P > 0
\end{equation}

If we approximate the opacity by a power law depending on density
and temperature $\kappa=C\rho^AT^B$, we can express the thermal
instability condition (\ref{eq:stabthermal}) for limiting cases in
terms of quantities depending on material functions only. We give
the exact derivation in Appendix \ref{sec:thermalviscous}. We
summarize the most useful formulae in Table \ref{tbl:thermal} (All
cases for $\mu=\mbox{const}$).

\begin{figure*}
\begin{center}
\epsfig{file=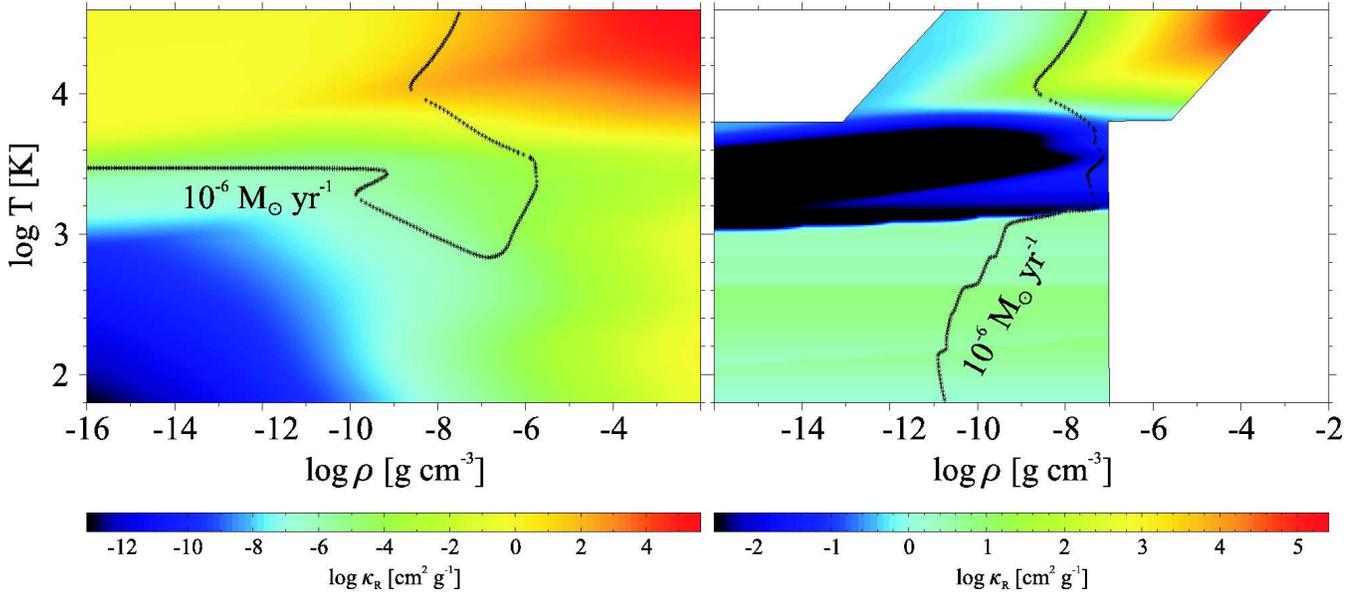,width=\textwidth}
\end{center}
\caption{Sample accretion disc model for Pop {\sc iii} (left,
Mayer \& Duschl, 2004, in prep.) and today's \citep[right,
][]{2003A+A...410..611S} chemical composition for $\dot M =
10^{-6} \Msun \yr^{-1}$. Note the different colouring schemes
mandated by the vastly different opacity range.}
\label{fig:disccomp}
\end{figure*}

\subsection{Viscous instability} \label{Sect:viscousinstability}

A disc is viscously unstable if an increase (decrease) of the
local accretion rate is answered by a decrease (increase) of the
surface density. Mathematically we can write this instability condition as
\begin{equation}
  \label{eq:stabvisc}
  \left.\left(\frac{\partial \log \dot M}{\partial \log \Sigma}\right)\right|_P < 0
\end{equation}

We call a disc thermal-viscously unstable if it is thermally and
viscously unstable.

For a power-law approximation of the opacity
$\kappa_{\Ross/\Planck}=C_0\rho^{A_{\Ross/\Planck}}T^{B_{\Ross/\Planck}}$
and $\mu=\mathrm{const}$ we get the results shown in Table
\ref{tbl:viscous}.

\begin{table}
\begin{center}
\renewcommand{\arraystretch}{1.7}

\begin{tabular}{|l|l|r|}
\hline \multirow{2}{1.3cm}{$P_{\gas}\gg P_{\rad}$} &
$\tau_{\Ross}\gg 1$ &
$\frac{3A_{\Ross}-(1+\eta_{\VSG})B_{\Ross}+5(1+\eta_{\VSG})}{A_{\Ross}-(1+\eta_{\VSG})B_{\Ross}+3(1+\eta_{\VSG})}<0$
\\\cline{2-3}
 & $\tau_{\Planck}\ll 1$ & $\frac{-3A_{\Planck}+(1+\eta_{\VSG})B_{\Planck}+3(1+\eta_{\VSG})}{-A_{\Planck}+(1+\eta_{\VSG})B_{\Planck}+3(1+\eta_{\VSG})}<0$ \\\cline{1-3}
\hline \multirow{2}{1.3cm}{$P_{\rad}\gg P_{\gas}$} &
$\tau_{\Ross}\gg 1$ &
$\frac{12A_{\Ross}+(2-\eta_{\VSG})B_{\Ross}-4(1-2\eta_{\VSG})}{4A_{\Ross}-\eta_{\VSG}
B_{\Ross}-4}<0$ \\\cline{2-3}
 & $\tau_{\Planck}\ll 1$ & $\frac{-12A_{\Planck}+(2\eta_{\VSG}-1)B_{\Planck}-4(2-\eta_{\VSG})}{-4A_{\Planck}+\eta_{\VSG} B_{\Planck}-4}<0$ \\\cline{1-3}
\hline
\end{tabular}

\end{center}
\caption{Conditions for viscous instability} \label{tbl:viscous}
\end{table}

\begin{table}
\begin{center}
\renewcommand{\arraystretch}{1.5}
\begin{tabular}{l|c|c}
\hline \hline
& $\left(\frac{\partial \log Q^+}{\partial \log T}\right)-\left(\frac{\partial \log Q^-}{\partial \log T}\right) =$ & $\left(\frac{\partial \log \dot M}{\partial \log \Sigma}\right) =$ \\
\hline
Ice grains & $(1+\eta_\VSG)$ & $3$ \\
\hline
Evaporation of ice gr. & $10(1+\eta_\VSG)$ & $\frac{6}{5}$\\
\hline
Metal grains & $\frac{5}{2}(1+\eta_\VSG)$ & $\frac{9}{5}$\\
\hline
Evaporation of metal gr. & $1+27(1+\eta_\VSG)$ & $\frac{3+29(1+\eta_\VSG)}{1+27(1+\eta_\VSG)}$\\
\hline
Molecules & $\frac{2}{3}$ & $6+3\eta_\VSG$\\
\hline
H-scattering & $\frac{1}{3}-7(1+\eta_\VSG)$ & $\frac{3-15(1+\eta_\VSG)}{1-21(1+\eta_\VSG)}$\\
\hline
Bound-free and free-free & $1+\frac{1}{2}(1+\eta_\VSG)$ &$\frac{6+5(1+\eta_\VSG)}{2+(1+\eta_\VSG)}$\\
\hline
Electron scattering ($\beta_P=0$)& $-4$ & $ 1-2\eta_\VSG$\\
\hline \hline
\end{tabular}
\renewcommand{\arraystretch}{1.0}
\end{center}
\caption{Stability criteria (eqs. \ref{eq:stabthermal} and
\ref{eq:stabvisc}) evaluated for the fits to Pop {\sc i} Rosseland
mean opacities from \citep{1994ApJ...427..987B}.}
\label{tbl:instab}
\end{table}

As an example we apply our instability criteria to the fits of
\cite{1994ApJ...427..987B} for Pop {\sc i} chemical composition
with $\tau_{\Ross} \gg 1$, assuming gas pressure to be dominant
($\beta_P=1$) except for Thomson scattering ($\beta_P=0$), and
show them in Table \ref{tbl:instab}.

We see a thermal instability due to H-scattering and a
thermal-viscous instability for electron scattering (for
$\eta_{\VSG}>1/2$ only) This is the so called
Lightman-Eardley instability \citep{1974ApJ...187L...1L}.

\section{Time scales}\label{Sect:timescale}

\subsection{Viscous, thermal and hydrostatical equilibrium}

The viscous, thermal and hydrostatic timescales are given by
\cite{1981ARA+A..19..137P}.
\begin{eqnarray*}
  \tau_{\mathrm{visc}}&=&\frac{r^2}{\nu}=\frac{r^2 \Omega}{\alpha c_\sound^2}\label{eq:timevisc}\\
  \tau_{\mathrm{th}}&=&\frac{\Sigma c_s^2}{Q^{+}}=\frac{\Sigma c_\sound^2}{\frac{9}{8}\nu \Sigma \Omega^2}=\frac{8}{9\alpha}\frac{1}{\Omega}\label{eq:timetherm}\\
   \tau_\ZZ&=& \frac{H}{c_\sound}\label{eq:timehor}
\end{eqnarray*}

\subsection{Chemical equilibrium}

A cautionary remark when dealing with Pop {\sc iii} matter should
be made regarding the chemical equilibrium. Due to the inefficient
cooling compared to dust (The primary coolant being $\Hmol$), the
only way to form $\Hmol$ in the low-density regime is via the
$\HH^-$ and $\Hmol^+$ channel
\citep[][]{1961Obser..81..240M,1968ApJ...154..891P,1967Natur..216..967S}.
However in the high-density regime ($\rho>10^{-16}\g \cm^{-3}$),
three-body reactions rapidly convert neutral atomic into molecular
hydrogen \citep{1983ApJ...271..632P}. A review describing chemical
reactions in the early Universe can be found in
\cite{2002JPhB...35R..57S}. We calculated the $\Hmol$ formation
timescale using the reaction rate $k_4$ of
\cite{1983ApJ...271..632P} (see Appendix~\ref{Sect:h2form}). We
get in the disc geometry
\begin{eqnarray*}
\tau_{\Hmol}&=& 0.51\cdot \xi(M,\rho,r)\cdot \left(\frac{\alpha}{0.01}\right)^{-\frac{2}{3}}\left(\frac{\dot M}{10^{-4}\Msun \yr^{-1}}\right)^\frac{2}{3}\\
&&\qquad \qquad \qquad \cdot \left(\frac{M}{\Msun}\right)^\frac{1}{3}\left(\frac{\rho}{10^{-8} \g \cm^{-3}}\right)^{-\frac{7}{3}}\left(\frac{r}{\AU}\right)^{-1}\left(\frac{\mu}{1}\right)^2 \s
\end{eqnarray*}
with $\xi(M,\rho,r)^3=1+4.76\left(\frac{M}{\Msun}\right)\left(\frac{\rho}{10^{-8} \g \cm^{-3}}\right)^{-1}\left(\frac{r}{\AU}\right)^{-3}$

\section{Population {\sc iii} accretion discs} \label{Sect:results}

\subsection{Overall Properties}

As we have seen in section \ref{Sect:opacity}, Pop {\sc iii}
matter gives rise to unusual absorption mechanisms like CIA. Due
to the lack of good absorbers it also cannot cool below a few 100
K \citep[for an overview of the initial conditions used for
collapse calculations see][]{2002MNRAS.334..401R} with the only
efficient coolant being mainly $\Hmol$. This minimum temperature
compared to the today's value around 10 K implies larger Jeans
masses \citep{2003MNRAS.343.1224C} and therefore larger accretion
rates. Some authors conclude therefore that Pop {\sc iii} stars
need to be more massive.

Fiducial values for primordial accretion discs are taken to be
$\dot M=10^{-4} \Msun \yr^{-1}$ compared to $\dot
M=10^{-6} \Msun \yr^{-1}$ for today's accretion
discs.

In our stationary accretion disc model the only parameters we can
adjust are the central mass, the accretion rate, and the viscosity
prescription. As standard value for $\alpha$ we use $\alpha=10^{-2}$.

\subsubsection{Self-similar solutions} \label{Sect:selfsimilar}

For all models presented there is no specification of an inner
boundary. In the scope of this paper, we want to keep the
discussion as general as possible.

Equations (\ref{eq:eq01}) and (\ref{eq:eq02}) are two equations
for the determinations of the density $\rho$ and the temperature
$T$ at a given radial distance $r$, central mass $M$ and accretion
rate $\dot M$. However $M$ and $r$ are coupled in the sense that
the equations yield the same result provided that
$\Omega^2=\frac{GM}{r^3}$, the rotation frequency is the same.
Thus we present our disc models in terms of a generalised radial
coordinate
\[
\log q = \log \left(\frac{r}{1\AU}\right)-\frac{1}{3}\log \left(\frac{M}{\Msun}\right)
\]

This approach only applies for $\NSG$ and $\VSG$ solutions. For
the FSG(full-selfgravitating) case $\Omega$ also depends on the
mass distribution of the disk itself.

For all models we choose a radial grid extending from
$10^{-3}\dots 10^3 \Msun^{-\frac{1}{3}}\AU $. At some accretion
rates the inferred radius of the central star can be larger than
our formal inner disc radius \citep[e.g. see Fig.1
in][]{2001ApJ...561L..55O}. The main scope of this paper is more
on the global properties of Pop {\sc iii} accretion discs in
comparison to Pop {\sc i} discs. Therefore we do not specify any
boundary conditions. This even makes the disk models applicable to
primordial BH accretion.

Sample accretion disc models are shown in Fig. \ref{fig:disccomp}.

\begin{figure*}
\rotatebox{90}{%
\begin{minipage}{\textheight}
\begin{center}
\epsfig{file=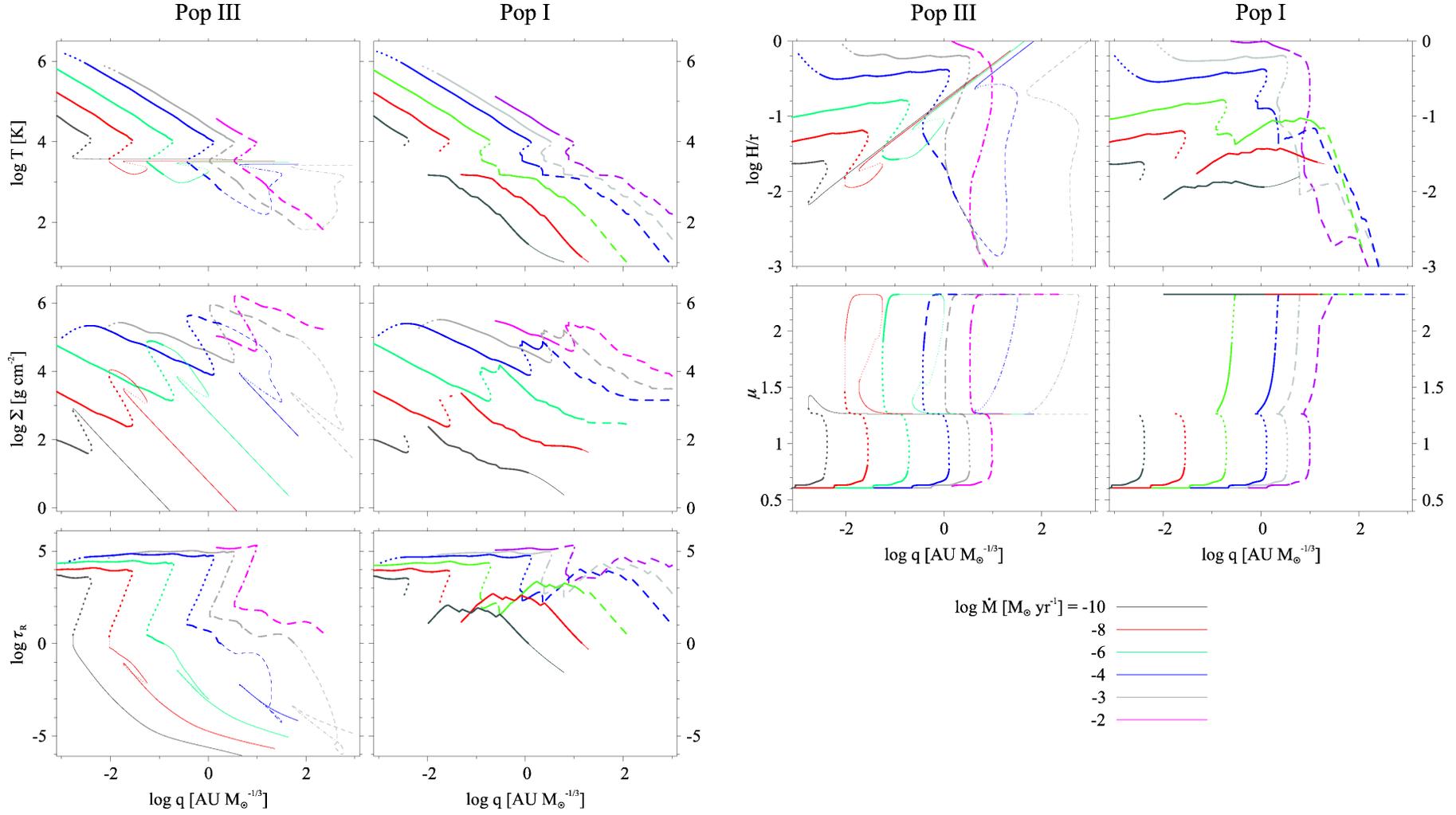,width=\textwidth}
\end{center}
\caption{Comparison of the properties of discs with primordial and
today's chemical composition: mid-plane temperature $T(r)$,
surface density $\Sigma(r)$, Rosseland optical depth $\tau_R(r)$,
relative thickness $H/r$, and mean molecular weight $\mu(r)$.
Solid lines are stable solutions, dotted lines correspond to
thermally unstable solutions and dashed lines are vertical
selfgravitating solutions. Dash-dotted lines are vertical
selfgravitating thermally unstable solutions.}
\label{fig:properties}
\end{minipage}%
}
\end{figure*}

\begin{figure*}
\epsfig{file=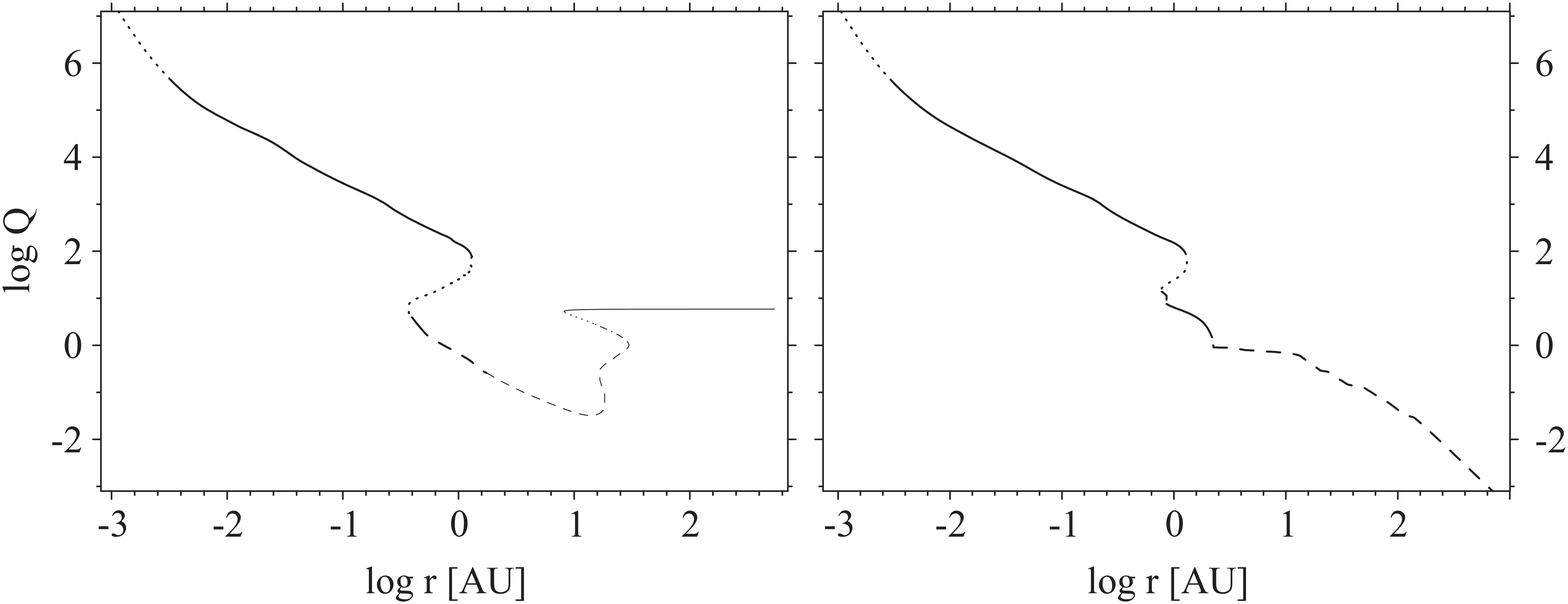,width=\textwidth}

\caption{Toomre parameter Q for a accretion rate of $10^{-4}
  \Msun \yr^{-1}$ in disc models of Pop {\sc iii} (left) and Pop {\sc
    i} (right) material. The disc is stable with respect to
  fragmentation for $\log Q>0$.} \label{fig:toomre}
\end{figure*}

\subsubsection{High-temperature limit} \label{Sect:hightemperature}

For high temperatures ($\log T\geq 3.8$) the models are nearly
independent of metallicity, because at high temperatures the opacity
is mainly dominated by hydrogen and helium species (up to an order of
magnitude). And due to the $T^4/\tau_{\eff}\propto \dot M \Omega^2$
proportionality a change in $\tau\propto \kappa_{\Ross}$ by one order
of magnitude changes the temperature only by a factor of 1.8 provided
$\Sigma$ remains constant.

\subsubsection{Isothermal outer disc} \label {Sect:isothermality}

One striking feature in the temperature $T(r)$ plot of Fig.
\ref{fig:properties} is the isothermal behaviour of the outer parts of
Pop {\sc iii} discs. The temperature seems to be locked to
approximately 3000 K. Inspecting the energy equation (\ref{eq:energy})
and replacing $\tau_{\eff}$ by $\frac{2}{3\tau_{\Planck}}$ with
$\tau_{\Planck}=\frac{1}{2}\Sigma
C_{\Planck}\left(\frac{\rho}{\rho_0}\right)^{A_{\Planck}}\left(\frac{T}{T_0}\right)^{B_{\Planck}}$,
we get:
\[
T=T_0\left(\frac{9 \alpha k }{8\sigma T_0^3 C_{\Planck}\mu
    m_p}\sqrt{\frac{GM}{r^3}}\left(\frac{\rho}{\rho_0}\right)^{-A_{\Planck}}\right)^{\frac{1}{B_{\Planck}+3}}
\]
For all absorption effects the exponent $A_\Planck$ remains smaller
than unity but the temperature dependence is far stronger. This is
because at temperatures of a few 1000 K, $\HH^-$ absorption is
dominant. In the optical thin part (where Planck mean values apply)
this absorption has a temperature dependence
$B_\Planck=\frac{\partial\log \kappa_{\Planck}}{\partial\log
T}\approx 20\dots 30$. The high value of $B_\Planck$ ensures that
then the exponent of the last equation is nearly zero and thus all
radial dependencies vanish. The disc is isothermal in radial
direction.

Such a temperature locking has also been shown for the spherical
collapse calculations of \cite{1986ApJ...302..590S} in the
photosphere of the accretion shock of an evolving primordial star.
Although the mechanism ($H^-$ absorption) is the same, the
isothermality for disk accretion is the property of the mid-plane
temperature, while for spherical accretion this is evident in the
photosphere of the evolving star.

Approximating the $\HH^-$ opacity (neglecting density dependence,
$A_{\Planck}=0$) we have
\[
\kappa_{\Planck,\HH^-}=C_{\Planck}\left(\frac{T}{3000 \mbox{
      K}}\right)^{B_{\Planck}}
\]
with $C_{\Planck}=10^{-8.4} \frac{\mbox{cm}^2}{\mbox{g}}$ and
$B_{\Planck}=22$.

Following our argumentation above we get
\[
T=3440 \mbox{K} \cdot
\left[\left(\frac{\alpha}{0.01}\right)\left(\frac{M}{\Msun}\right)^\frac{1}{2}\left(\frac{r}{\mbox{AU}}\right)^{-\frac{3}{2}}\right]^\frac{1}{3+B_{\Planck}}
\]

In today's accretion discs at 3000 K molecular opacities of
$\Hmol\mathrm{O}$ and $\mathrm{TiO}$ soften the temperature dependence
and therefore let the temperature decrease significantly. With
decreasing temperature we get into the region of dust opacity which
does not provide these steep temperature gradients.

\subsubsection{Disc thickness} \label{Sect:hr}

As one consequence of this radial isothermality in Pop {\sc iii} discs
the relative disc thickness $H/R$ increases as the disc is optically
thin and $H/R=c_s/\left(\Omega
  r\right)\left(1+\eta_\VSG/\left(1-\eta_\VSG\right)\right)^\frac{1}{2}\propto
c_s/\left(\Omega r\right)\propto \left(Tr\right)^\frac{1}{2}$. For
constant $T$ we have $H/R \propto r^\frac{1}{2}$. This leads to the
break-down of the thin-disc approximation as $H/R$ reaches and exceeds
unity. The other consequence is the radial decline of the surface
density $\Sigma \propto r^{-\frac{3}{2}}$. We use the angular momentum
equation (\ref{eq:angmom}) and get $c_\sound^2 \Sigma \propto \Omega
\dot M$ and with isothermality $\Sigma \propto \Omega\propto
r^{-\frac{3}{2}}$.

In the dust opacity disc models the temperature is not constant in the
outer disc. Therefore $H/R$ decreases with the onset of vertical
selfgravity, as then $H/R\propto T/\left(\Sigma r\right)\approx T/r$.

\subsubsection{Minimal temperature}

Between the hot inner part and the isothermal outer part, stationary
Pop {\sc iii} discs have an intermediate region, where the disc can
cool below the 3000 K of the outer parts. This is mainly due to the
onset of CIA absorption. But as the density decreases, CIA absorption
gets less and less efficient (absorption coefficient proportional to
the product of number densities of the contributing absorbers). At a
certain point the absorption is so inefficient that the disc becomes
optically thin and the absorption turns into optically thin emission.
The temperature has to rise again until it reaches that of the
temperature locked outer region. The minimal temperature decreases
with increasing accretion rate as then more and more material is
accreted. Therefore the densities get higher and so does absorption.

\subsection{Instabilities}

\subsubsection{Gravitational instability} \label{Sect:toomre}

Figure \ref{fig:toomre} shows the Toomre parameter for Pop {\sc iii}
and Pop {\sc i} discs. While the Toomre parameter decreases
monotonically for Pop {\sc i} discs (as temperature does) there is a
difference in the Pop {\sc iii} case.  The temperature reaches a
minimum value and so does the Toomre parameter.

With the analytical considerations of Sect. \ref{Sect:hr} we can write
the Toomre parameter (\ref{eq:toomre}) for the optically thin
isothermal outer region as
\[
Q= 10.7 \left(\frac{\alpha}{0.01}\right)\left(\frac{\dot
    M}{10^{-4}\Msun
    \yr^{-1}}\right)^{-1}\left(\frac{T}{3440\mbox{K}}\right)^\frac{3}{2}
\]

For accretion rates in excess of $1.07\cdot 10^{-3} \Msun \yr^{-1}$
($\alpha=0.01$) the necessary condition for fragmentation is fulfilled
in the Pop {\sc iii} case.

In isothermal regions the Toomre parameter depends only on radially
constant quantities. The surface density $\Sigma$ and the rotation
frequency $\Omega$ both show the same radial dependence which
cancels in eq. (\ref{eq:toomre}).


\revs{The maximum value of the viscosity parameter for marginally
stable, selfgravitating discs is still a matter of debate:
\citet{2004ApJ...603..401T} use $\alpha=0.3$, following
\citet{2003MNRAS.339..937G}, while \citet{2001ApJ...553..174G} finds
analytically $\alpha\approx \frac{2}{90}\approx 0.022$ and
numerically $\alpha\approx 0.0247$ (see his Sect. 4.1). By using
$\alpha=0.01$ we underestimate his critical accretion rate for
fragmentation by a factor of order unity. Given the fact that the
true critical Toomre parameter is still under debate (cf. Sect.
\ref{sect:toomre}), using $\alpha=0.01$ seems to us to be adequate.
Using $\alpha=0.3$ for marginal selfgravity, as Tan \& Blackman do,
rises the critical accretion rate by a about one and a half orders
of magnitude.}

The role of selfgravity in Pop {\sc iii} is different compared to
those found in Pop {\sc i}. Usual low-temperature dust dominated
Pop {\sc i} discs become vertically selfgravitating (KSG) as the
disc mass exceeds the fraction $\frac{h}{r}$ of the accreting mass
and become fully selfgravitating (FSG) for disc masses larger than
the accreting mass while remaining vertical selfgravitating
\citep{2000A&A...357.1123D}. However, in the Pop {\sc iii} case
the outer parts can be FSG while not any more being KSG due to the
local mass density.

This can be understood by examining the ratio
\[
\frac{g_{\VSG}}{g_{c,z}}=\frac{4\pi
  \rho r^3}{M}
\]
For the Pop {\sc i} case, due to cooling we have the decrease of the
disc $H/R$ ratio ($H/R\propto R^{-\xi}$, $\xi\approx 2$). With
assuming the surface density $\Sigma\propto R^{-\zeta}$, $\zeta\approx
0$), the density is $\rho\propto \Sigma/H\propto R^{\xi-\zeta-1}$. The
ratio $g_{\VSG}/g_{c,z}\propto R^{\xi-\eta+2}$ increases with radius
provided $\xi-\zeta>-2$ which is mostly satisfied in the Pop {\sc i}
case.

The scaling in Pop {\sc iii} is different. $H/R\propto R$ and
$\Sigma\propto R^{-\frac{3}{2}}$ (see Sect.~\ref{Sect:hr}). Hence
$\rho \propto \Sigma/H\propto R^{-\frac{7}{2}}$. Once the disc reaches
the temperature-locked region (Sect.~\ref{Sect:isothermality}),
$g_\VSG/g_\CC\propto r^{-\frac{1}{2}}$. The vertical selfgravity is
diminishing, while the disc mass in the isothermal part still
increases. ($M_{\rm d,isothermal}=\int 2\pi \Sigma r dr\propto
r^{\frac{1}{2}}$).

Thus in the presence of Pop {\sc iii} matter one needs to distinguish
between the local $\VSG$-criterion ($g_\VSG/g_\CC\ge 1$) and the
global FSG-criterion $M_d \ge M$.

\subsubsection{Thermal and viscous instabilities}

\begin{figure*}
\epsfig{file=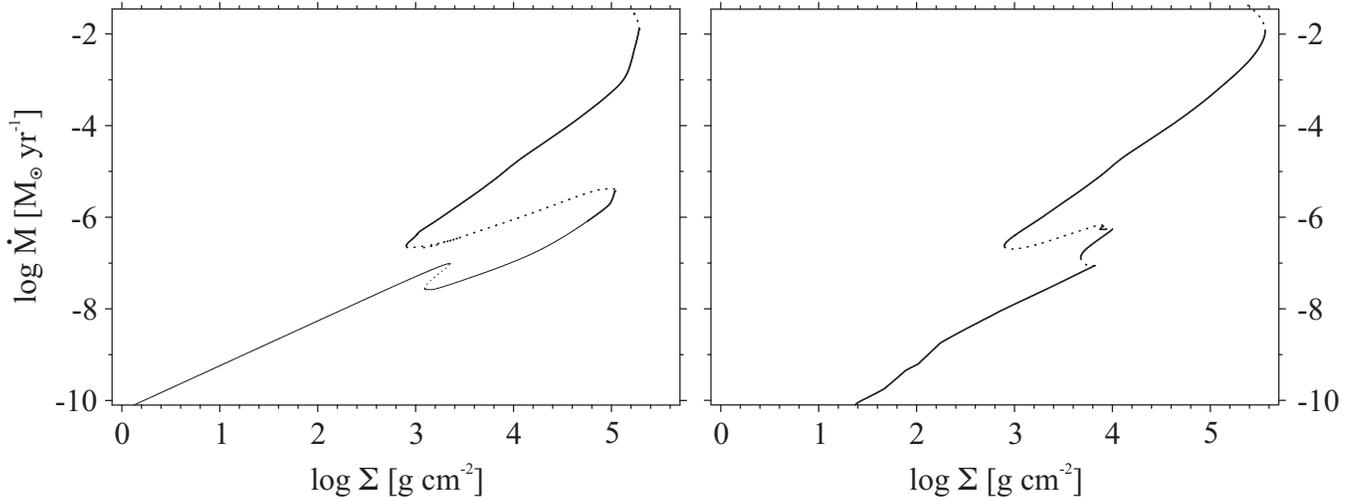,width=\textwidth}

\caption{Accretion rate versus surface density at radial distance
  of $r=0.1$ AU in disc models of Pop {\sc iii} (left) and Pop {\sc i}
  (right) material with central mass of $1\Msun$. The meaning of the lines, dots, etc. is the same
  as Fig. \ref{fig:properties}.} \label{fig:msigma}
\end{figure*}

In Fig. \ref{fig:properties} we already indicate thermal
instabilities. We now want to examine the effect of these
instabilities by varying the accretion rate $\dot M(\Sigma)$ at
given radial distance as shown in Fig. \ref{fig:msigma}. According
to eq. (\ref{eq:stabvisc}) the disc is viscously unstable for
negative $\dot M$-$\Sigma$ slope. This is the case at very high
accretion rates (Lightman-Eardley instability) in both chemical
compositions and also at some kinks in the molecular opacities of
Pop {\sc i} material. In our Pop {\sc iii} disc models we do not
find any viscous instabilities except at the turning points
between the thermally stable and unstable solutions.

Fig. \ref{fig:msigma} shows that limit cycles are present. For both
POP {\sc i} and POP {\sc iii} matter the limit cycles are mainly due
to $\mbox{H}^-$-absorption. The influence of molecular lines (TiO,
CO, etc.), however, causes further kinks at lower accretion rates
and prevents Pop {\sc i} disks from undergoing large $\Delta \dot M$
- $\Delta \Sigma$ outbursts. For POP {\sc iii} no such molecular
lines are present at these temperatures ($\mbox{H}_2$ lines set in
at lower temperatures). At the transition from the intermediate CIA
cooled region of the disc to the isothermal outer disc there is an
additional thermal instability present. This instability arises
because at this temperatures $\mbox{H}_2$ dissociates while the
contribution of atomic hydrogen to the opacity is not yet that
large.

\subsection{Timescales} \label{Sect:h2formres}

\begin{figure*}
\epsfig{file=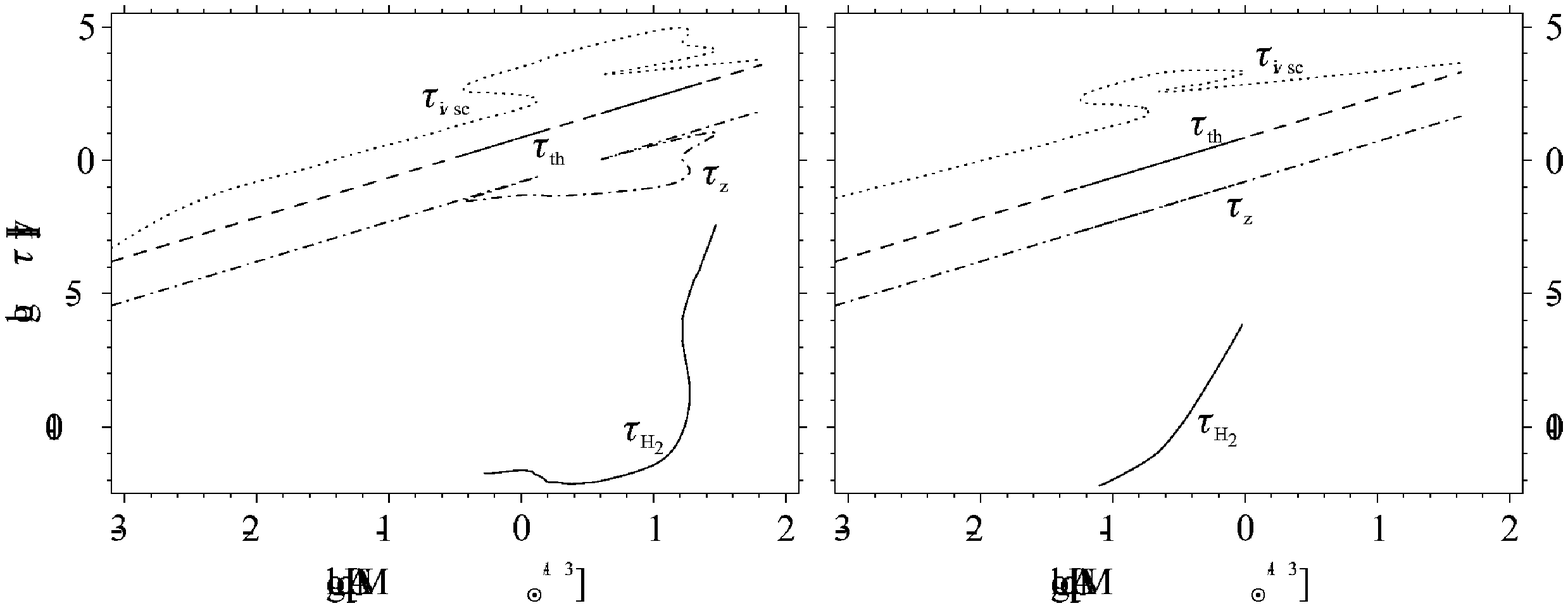,width=\textwidth}
  \caption{Timescales for Pop {\sc iii} disc models with accretion rates $10^{-4}\mbox{ (left) and }10^{-6} \mbox{ (right) }\Msun\yr^{-1}$.}\label{fig:Timescale}
\end{figure*}

For the assumptions in our disc model (hydrostatic, thermal and
chemical equilibrium, 1+1D one-zone approximation) we need
\[
\tau_{\mathrm{visc}}\gg\tau_{\mathrm{th}}\gg\tau_{\ZZ}\gg \tau_{\Hmol}
\]
to be satisfied. We show the timescales for sample models in Fig.
\ref{fig:Timescale}. The chemical equilibrium is implicitly
assumed to be fullfilled everywhere except where we have a
significant abundance of $\Hmol$. In this domain we plot the
$\Hmol$ formation timescale. Otherwise there are only ionic and/or
ion-neutral chemical rections which proceed on timescales much
smaller than timescales present in the disc.

The assumption of
chemical equilibrium is satisfied even in the Pop {\sc iii} case. The
$\Hmol$ formation timescale is orders of magnitudes smaller than the
vertical timescale except in the outer regions of the disc. As long as
the density is high enough, the 3-body reactions rapidly convert
atomic into molecular hydrogen. In the outer parts of the disc, where
$\tau_{\Hmol}$ increases, the strong density dependence of the 3-body
reaction time scale can be seen.

However, our $\Hmol$ formation timescale should only be used with care.
Our assumption is that this timescale equals the chemical reaction
timescale. In general we only could assume that $\tau_{\Hmol}$ is a
lower limit to the chemical reaction timescale as the consumed $\HH$
still has to be depleted to reach chemical equilibrium.

As, however, $\tau_{\Hmol}$ is several orders of magnitude smaller
than the hydrostativ timescale, atomic hydrogen has got time
enough to be sufficiently depleted not influencing the opacity and
therefore the disc structure. In the low-temperature region the
only important absorption mechanism influencing the opacity is
CIA.

The global maximum of the absorption coefficient normalized to
standard number densities approximately coincide at the same level
and the same frequency. Due to the quadratic dependence of the CIA
on the density of the contributing species, once $\Hmol$ has been
formed, CIA of $\Hmol/\Hmol$ pairs is the dominant absorption
mechanism at high densities.

\section{Discussion} \label{Sect:discussion}

We have shown the impact of the different chemical composition on
stationary accretion discs, comparing Pop {\sc iii} and Pop {\sc i}
material. This is mainly due to the lack of dust and heavier molecules
($\Hmol\mathrm{O}$, $\mathrm{TiO}$, etc.) absorption at low
temperatures.

While the inner (hot) parts of Pop {\sc iii} discs do not largely
differ from Pop {\sc i} discs due to the dominant influence of
hydrogen and helium species on the absorption, Pop {\sc iii} are
optically thin and isothermal in their outer parts because of
temperature locking due to $\mbox{H}^-$-absorption in the absence of
dust. This leads to the breakdown of the thin-disc approximation as
$H/R$ increases proportional to $r^\frac{1}{2}$.  Pop {\sc i} discs do
not have this locking mechanism. Therefore $H/R$ decreases with the
onset of the vertical selfgravity. This considerable flaring may lead
to a disc structure modified by irradiation from the accreting star
and the inner disc regions.  Such effects, however, are beyond the
scope of this paper.

The disk models considered in this paper do not take into account
effects due to the inner or outer boundary condition. Thus for
protostellar evolution the innermost solutions will be nonexistent
being part of the protostar, while the outer disk regions may be
hidden in spherical accretion still ongoing
\citepalias[see][]{2004ApJ...603..383T} .

At the very outermost parts of Pop {\sc iii} accretion discs the
thermal timescale becomes larger than the viscous timescale as
$H/R$ reaches unity (see Fig. \ref{fig:Timescale}). However it
remains to be shown wether the breakdown of the thin-disk
approximation (Sect.\ \ref{Sect:hr}) remains valid when taking
into account irradiation.

The assumption of chemical equilibrium is well satisfied for all
regions with reasonable disc solutions in contrast to primordial
protostellar collapse models. Given the much smaller density, the
$\Hmol$ formation timescale exceeds the dynmical time scale there.

Pop {\sc iii} discs have intermediate regions, where they are able
to cool below the 3440 K of the outer temperature-locked zones,
where CIA is the dominant absorption effect. The minimal
temperature achievable decreases with increasing accretion rate.
For accretion rates high enough these discs do have a vertically
selfgravitating ring (see Fig. \ref{fig:properties}).

This selfgravitating ring may be subject to change after inclusion
of line absorption in the opacity calculations. At low density and
low temperature we expect primarily $H_2$ quadrupole line
absorption which may prevent the disc from reaching the isothermal
outer part. As the molecular lines are narrow and weak at low
densities, the integrated absorption coefficient added to the
continuum Planck opacities will become constant with decreasing
density for given temperature($\kappa_P(\rho,T)\propto T^A$),
while it does not in the continuum case ($\kappa_P(\rho,T)\propto
\rho T^B$). In this line absorption dominated regime, the Pop {\sc
iii} opacities may exhibit the same behaviour as Pop {\sc i}
opacities do in their dust and ice dominated regime, although at
much lower absolute values.

In this case, there may be at least two coexisting solutions: The
isothermal part will remain (temperature is only fixed due to
$H^-$ absorption which will remain the dominant absorption in this
temperature range) with the addition of a stable low-temperature
solution branch.

As mentioned above, some authors conclude that Pop {\sc iii} stars
must have been very massive. There have been many proposals of
feedback phenomena which may halt the accretion
\citep{2001ApJ...561L..55O, 2003ApJ...589..677O}. Our stability
examinations show a direct limiting process. While with Pop {\sc
i} discs the accretion rate due to the thermal instability can be
increased by 1.5 orders of magnitude in contrast to nearly 3
orders of magnitude for Pop {\sc iii} discs, Pop {\sc iii} discs
are likely to exceed their Eddington Limit quite often and
therefore produce massive outflows. However it remains to be shown
in time-dependent simulations of Pop {\sc iii} discs how this
instability acts.

The Toomre parameter of the isothermal outer region becomes lower
than unity if the accretion rate exceeds $1.07\cdot 10^{-3} \Msun
\yr^{-1}$ (Sect. \ref{Sect:toomre}). \revs{While
\citet{2004ApJ...603..401T} use $\alpha=0.3$ (following
\citet{2003MNRAS.339..937G}) we use $\alpha=0.01$ as the value for
the viscosity parameter following \citet{2001ApJ...553..174G}.}

Discs well above this threshold accretion rate are certainly prone
to fragmentation. Discs with lower rates can have Toomre parameter
values $Q<1$ in the intermediate regime but $Q>1$ in the
isothermal outer parts. This does not lead to framentation, as the
viscous timescale in the outer parts of Pop {\sc iii} accretion
disc decreases until the onset of the isothermal part. This means
that the first stars can grow very large and that it is
exceedingly hard to form Pop {\sc iii} planets.

\section*{Acknowledgements}
The authors acknowledge support for this work from the {\it
Deut\-sche For\-schungs\-ge\-mein\-schaft, DFG\/} through grant
{\it SFB 439 (A7)\/}. Part of this work was carried out while one
of the authors (MM) stayed as an EARA Marie Curie Fellow at the
Institute of Astronomy, University of Cambridge, UK. The
hospitality of the IoA, and of Prof. J.E. Pringle in particular,
is gratefully acknowledged. The authors thank Drs. Christian
Straka and Franck Hersant for valuable discussions on the subject
of this paper. Finally we thank the referee, Dr. Jonathan Tan for
his comments on the manuscript, which helped to improve it.


\appendix

\section{The implicit equation for the viscosity} \label{sect:viscositydet}

In the case of the $\alpha$-viscosity the viscosity can be written as
a function of density $\rho$ and temperature $T$. We have
\begin{eqnarray*}
  \nu&=&\alpha\frac{P}{\Omega\rho}\\
  &=& \frac{\alpha}{\Omega} \left[\frac{kT}{\mu m_{\mbox{p}}}+\frac{4\sigma T^4}{3c\rho\tau_{\eff}(\nu)}\left(\frac{\kappa_{\Ross}\dot M}{6\pi \nu}+\frac{4}{3}\right)\right]
\end{eqnarray*}
making use of the angular momentum equation and
\begin{eqnarray*}
  \tau_{\eff}&=&\frac{\kappa_{\Ross}\dot M}{6\pi \nu}+\frac{4}{3}+\frac{4\pi \nu}{\kappa_{\Planck} \dot M}\\
&=& \frac{T_1+T_2\nu+T_3\nu^2}{\nu}
\end{eqnarray*}
with
\[
T_1=\frac{\kappa_{\Ross}\dot M}{6\pi}\qquad T_2=\frac{4}{3}\qquad
T_3=\frac{4\pi}{\kappa_{\Planck}\dot M}
\]
and further setting
\[
C_{\gas}=\frac{kT}{\mu m_{\mbox{p}}}\qquad C_{\rad}=\frac{4\sigma
  T^4}{3c\rho}\qquad C_0=\frac{\alpha}{\Omega}
\]
we get
\begin{eqnarray*}
  \nu&=& C_0 \left[C_{\gas}+C_{\rad}\frac{T_1+T_2\nu}{T_1+T_2\nu+T_3\nu^2}\right]
\end{eqnarray*}
\begin{eqnarray*}
  T_3\nu^3 + \left(T_2-C_0C_{\gas}T_3\right)\nu^2+\left(T_1-C_0T_2\left(C_{\gas}+C_{\rad}\right)\right)\nu &&\\
  -C_0T_1\left(C_{\gas}+C_{\rad}\right) &=&0
\end{eqnarray*}

This equation has to be solved for given density and temperature.

\section{Thermal and viscous stability of radiatively supported discs} \label{sec:thermalviscous}

In this section, variables like $\bar x$ denote logarithmic variables.

We start with the equations for energy and hydrostatic equilibrium
(see. eqns. \ref{eq:energy} and \ref{eq:pressure})
\begin{eqnarray*}
  \frac{9}{8}\nu \Sigma\Omega^2&=&\frac{4\sigma}{3c\tau_{\eff}}T^4\\
  \rho \frac{kT}{\mu m_{\mbox{p}}}+\frac{4\sigma}{3c\tau_{\eff}}T^4\left(\tau_{\Ross}+\frac{4}{3}\right)&=& G\left(\pi+\frac{M}{4\rho r^3}\right)\Sigma^2
\end{eqnarray*}

Using eqns. (\ref{eq:pressure2}), (\ref{eq:tau}) and
(\ref{eq:rossplanck}) we can write $\rho$ and $T$ as a function of
$\Sigma$. $\nu$ and $\tau_{\eff}$ remain functions of $\rho$, $T$ and
$\Sigma$.
\begin{equation}
  \label{eq:relations}
  \rho(\Sigma) \quad T(\Sigma)\quad \tau_{\eff}(\rho,T,\Sigma)\quad \nu(\rho,T,\Sigma)
\end{equation}

\subsection{Viscous instability criterion}
We are interested in the derivative
\begin{equation}
  \label{eq:accratederiv}
  \left(\frac{\partial \dot {\bar M}}{\partial \bar \Sigma}\right)_{d\bar Q^+=d\bar Q^-,d\bar P=0}= \frac{\partial \dot{\bar M}}{\partial \bar \rho}\frac{\partial \bar \rho}{\partial \bar \Sigma}+\frac{\partial \dot {\bar M}}{\partial \bar T}\frac{\partial \bar T}{\partial \bar \Sigma}+\frac{\partial \dot{\bar M}}{\partial \bar \Sigma}
\end{equation}

This derivative has to be calculated for equilibrium pressure and
energy. Thus we have
\begin{eqnarray*}
  d\left(\bar Q^+-\bar Q^-\right)&=&0\\
  d\left(\overline{P_{\gas}+P_{\rad}}-\overline{g_{\CC}\Sigma+g_{\VSG}\Sigma}\right)&=&0
\end{eqnarray*}
which results in 2 linear equations for the unknown derivatives
$\frac{\partial \bar \rho}{\partial \bar \Sigma}$ and $\frac{\partial
  \bar T}{\partial \bar \Sigma}$
\begin{eqnarray*}
  A \frac{\partial \bar \rho}{\partial \bar \Sigma} + B \frac{\partial \bar T}{\partial \bar \Sigma} + C &=& 0\\
  D \frac{\partial \bar \rho}{\partial \bar \Sigma} + E \frac{\partial \bar T}{\partial \bar \Sigma} + F &=& 0
\end{eqnarray*}
which can be calculated for $AE-BD\ne 0$ to
\[
\frac{\partial \bar \rho}{\partial \bar \Sigma}=\frac{BF-CE}{AE-BD}
\qquad \frac{\partial \bar T}{\partial \bar
  \Sigma}=\frac{CD-AF}{AE-BD}
\]

With a further examination of eq. (\ref{eq:accratederiv}) we get
\begin{eqnarray*}
\left.\left(\frac{\partial \dot {\bar M}}{\partial \bar \Sigma}\right)\right|_P&=& \frac{\partial \dot{\bar M}}{\partial \bar \rho}\frac{\partial \bar \rho}{\partial \bar \Sigma}+\frac{\partial \dot {\bar M}}{\partial \bar T}\frac{\partial \bar T}{\partial \bar \Sigma}+\frac{\partial \dot{\bar M}}{\partial \bar \Sigma}\\
&=& \frac{\partial \bar \nu}{\partial \bar \rho}\frac{\partial\bar \rho}{\partial \bar \Sigma}+\frac{\partial \bar \nu}{\partial \bar T}\frac{\partial \bar T}{\partial \bar \Sigma}+\frac{\partial \bar \nu}{\partial \bar \Sigma}+1\\
&=& \frac{f\left(A,B,C,D,E,F,\frac{\partial\bar\nu}{\partial\left(\bar \rho,\bar T,\bar \Sigma\right)}\right)}{AE-BD}
\end{eqnarray*}
\begin{equation}
  \label{eq:viscousinstability}
  \left.\left(\frac{\partial \dot {\bar M}}{\partial \bar \Sigma}\right)\right|_P=\frac{f\left(A,B,C,D,E,F,\frac{\partial\bar\nu}{\partial\left(\bar \rho,\bar T,\bar \Sigma\right)}\right)}{AE-BD}<0
\end{equation}
\subsection{Thermal instability criterion}

Thermal instability requires

\begin{equation}
  \label{eq:thermal}
  \left.\left(\frac{\partial \bar Q^+}{\partial \bar T}\right)\right|_P-\left.\left(\frac{\partial \bar Q^-}{\partial \bar T}\right)\right|_P>0
\end{equation}
to be satisfied. We expand $Q^+$ and $Q^-$ in terms of $\rho$ and $T$
\begin{eqnarray*}
d\left(\bar Q^+-\bar Q^-\right) &=& \left(\underbrace{\frac{\partial\bar Q^+}{\partial\bar \rho}-\frac{\partial\bar Q^-}{\partial\bar \rho}}_{=A}\right)d\bar\rho+\left(\underbrace{\frac{\partial\bar Q^+}{\partial\bar T}-\frac{\partial\bar Q^-}{\partial\bar T}}_{=B}\right)d\bar T\\
&=& Ad\bar\rho+Bd\bar T
\end{eqnarray*}
We further need to assure $d\bar P(\bar \rho,\bar T)=0$
\[
d\left(\overline{P_{\CC}+P_{\VSG}}-\overline{P_{\gas}+P_{\rad}}\right)=Dd\bar
\rho+Ed\bar T=0
\]
This is a condition which expresses $d\bar \rho$ in terms of $d\bar
T$.  We get our thermal instability condition according to
eq.~(\ref{eq:thermal}) to
\[
\left.\left(\frac{\partial \bar Q^+}{\partial \bar
      T}\right)\right|_P-\left.\left(\frac{\partial \bar Q^-}{\partial
      \bar T}\right)\right|_P=-\frac{AE}{D}+B>0
\]
As $D=\frac{\partial \bar P}{\partial\bar \rho}\ge 0$
\begin{equation}
AE-BD<0 \label{eq:thermalinstability}
\end{equation}
The term $AE-BD$ is the same as in the denominator of the viscous
instability criterion. Therefore it can be conjectured that if the
disc is viscously unstable there also may be a thermal instability
present. This leads to the expression of thermal-viscous instability.
However this relation does not hold in a strict sense. There are at
least formally possibilities having viscous instabilities without
having thermal instabilities.

The conditions (\ref{eq:thermalinstability}) and
(\ref{eq:viscousinstability}) can in principle be expressed now
depending only on $A_{\Ross/\Planck}$, $B_{\Ross/\Planck}$, $\beta_P$
and $\eta_{\VSG}$ with contributions of the $\mu$-gradients in density
and temperature. However we restrict ourselves only to supply handy
formulae for relevant limit cases.  (See Section
\ref{Sect:thermalinstability} and \ref{Sect:viscousinstability})

\section{Estimation of the $\Hmol$ formation time scale} \label{Sect:h2form}

We calculate the disc properties analytically for a selfgravitating
ring, assuming $\beta_P=1$.

\begin{figure*}

\centering

\epsfig{file=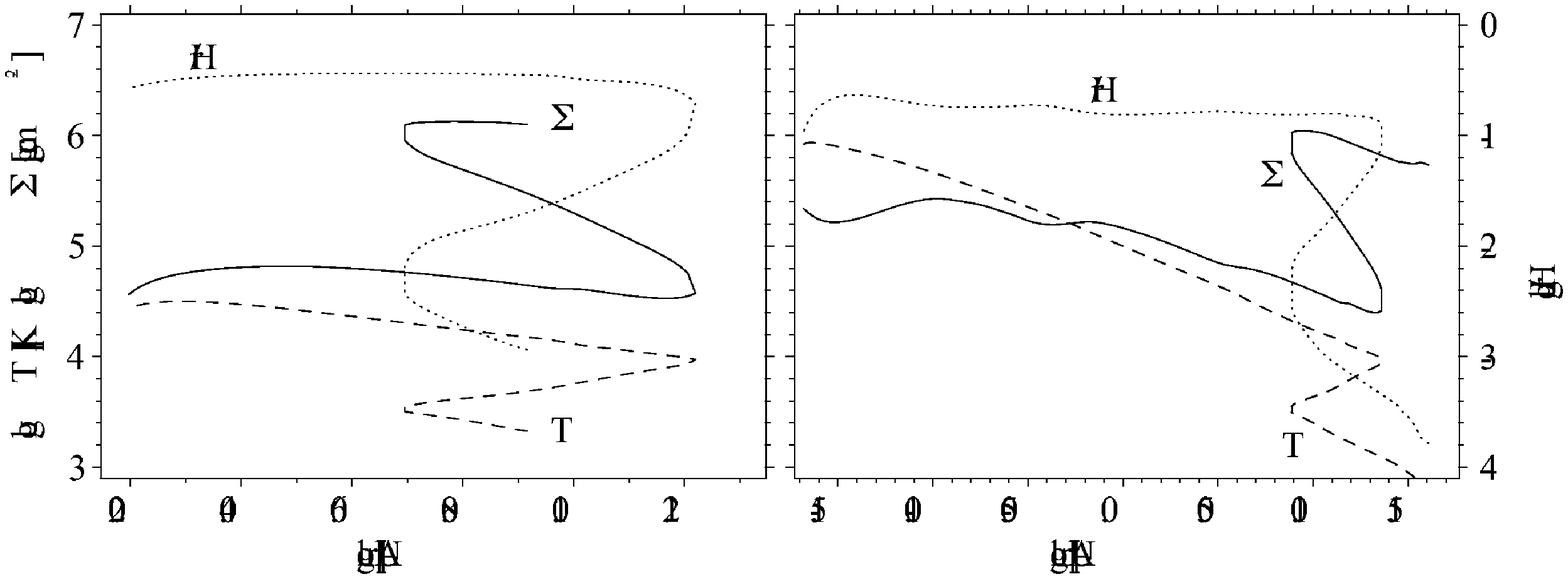,width=\textwidth}

\caption{Accretion disc models for two cases of Tan \& McKee
(2004): $M=10\Msun$, $r_*=300 R_\odot$, $\dot M=6.4\cdot 10^{-3}
\Msun \yr^{-1}$ (left), $M=100\Msun$, $r_*=4 R_\odot$, $\dot
M=2.4\cdot 10^{-3} \Msun \yr^{-1}$ (right). $\alpha$ was chosen
0.01. We discard all solutions where $\Sigma s^2$, an
approximation for the disc mass exceeds the central mass.}
\label{fig:tankee}
\end{figure*}

>From the angular momentum equation (\ref{eq:angmom}) with the
our viscosity prescription (\ref{eq:alphaP}) we get
\[
\dot M = 3\pi \alpha \frac{kT}{\mu m_p}\Sigma \sqrt{\frac{r^3}{GM}}.
\]
This leads to
\[
\Sigma(T)= \frac{\dot M \mu m_p}{3\pi \alpha k T}\sqrt{\frac{GM}{r^3}}.
\]
Plugging this into the equation for the hydrostatic equilibrium leads to
(\ref{eq:pressure})
\[
\rho\frac{kT}{\mu m_p}=G\left(\frac{M}{4\rho r^3}+\pi\right)\frac{\dot
  M^2 \mu^2 m_p^2}{9\pi^2 \alpha^2 k^2 T^2}\frac{GM}{r^3}.
\]
This is an equation which can be solved for $T(\rho)$.
\begin{eqnarray*}
T^3&=&\left(\frac{M}{4\rho r^3}+\pi\right)\frac{\dot M^2 \mu^3 m_p^3}{9\pi^2 \alpha^2 k^3\rho}\frac{G^2M}{r^3}\\
T&=& 2010 \mbox{ K }\cdot  \xi(M,\rho,r)\cdot \left(\frac{\alpha}{0.01}\right)^{-\frac{2}{3}}\left(\frac{\dot M}{10^{-4}\Msun \yr^{-1}}\right)^\frac{2}{3} \\
&&\qquad \qquad \qquad \cdot \left(\frac{M}{\Msun}\right)^\frac{1}{3}\left(\frac{\rho}{10^{-8} \g \cm^{-3}}\right)^{-\frac{1}{3}}\left(\frac{r}{\AU}\right)^{-1}\left(\frac{\mu}{1}\right)^1\\
\end{eqnarray*}
with
\[
\xi(M,\rho,r)=\left(1+4.76\left(\frac{M}{\Msun}\right)\left(\frac{\rho}{10^{-8}
      \g
      \cm^{-3}}\right)^{-1}\left(\frac{r}{\AU}\right)^{-3}\right)^\frac{1}{3}
\]
With $k_4=1.83\cdot 10^{-31}\left(\frac{T}{300 \mathrm{ K}}\right)^{-1} \cm^6
\s^{-1}$ \citep{1983ApJ...271..632P} and $\rho=\mu m_p n_H$ we get for
the H$_2$ formation time scale:
\begin{eqnarray*}
\tau_{\mathrm{H}_2}&=&\frac{1}{2k_4n_H^2}\\
&=&  7.65 \cdot 10^{-2}\s \cdot \left(\frac{T}{300 \K}\right) \left(\frac{\rho}{10^{-8} \g\cm^{-3}}\right)^{-2} \left(\frac{\mu}{1}\right)\\
&=& 0.51 \s \cdot \xi(M,\rho,r)\cdot \left(\frac{\alpha}{0.01}\right)^{-\frac{2}{3}}\left(\frac{\dot M}{10^{-4}\Msun \yr^{-1}}\right)^\frac{2}{3}\\
&&\qquad \qquad \qquad \cdot \left(\frac{M}{\Msun}\right)^\frac{1}{3}\left(\frac{\rho}{10^{-8} \g \cm^{-3}}\right)^{-\frac{7}{3}}\left(\frac{r}{\AU}\right)^{-1}\left(\frac{\mu}{1}\right)^2.\\
\end{eqnarray*}
Note : For $\xi=1$ we are in the vertical selfgravitating domain, $\xi\ge 1$
elsewhere.

Alternatively we could have taken the reaction rate
$k_{3b}=1.3\cdot 10^{-32}\left(\frac{T}{300 \K}\right)^{-1} \cm^6
\s^{-1}$ of \cite{2002Sci...295...93A}. Then the corresponding
numerical factors would be 1.07 and 7.23. Using $k_{3b}$ instead
of $k_4$ leads to an order of magnitude higher value of
$\tau_{\Hmol}$.

However, with more and more $H_2$ being produced, another 3-body
reaction \citep[$k_6$ of][]{1983ApJ...271..632P} will help in
further reducing the formation time scale.

\section{Comparison to the Tan \& McKee (2004) disc models} \label{Sect:tmc}

\citetalias{2004ApJ...603..383T} present disc models for the inner
regions of the accretion disc around a Pop {\sc iii} star. Their
disc models span 1 decade in radial distance (with the exception
of the 100 $\Msun$ case), are strongly influenced by the inner
boundary condition. In order to compare to our model, we subjected
our discs to the same inner boundary. We present our results for
two of their parameter sets in Fig. \ref{fig:tankee}.

The results are very similar at the high-temperature region.
However, the \citetalias{2004ApJ...603..383T} neglect the
low-temperature region. Our models show that there are two stable
solutions coexisting over large radial distances, connected
through the thermally unstable part (cf. Fig.
\ref{fig:properties}), giving rise to thermal instabilities.

\citetalias{2004ApJ...603..383T} further mention the importance of
the ionisation term in the energy equation, while the thermal term
is negligible. We do not account for this effect, as we want to
keep the presentation as general as possible (see Sect.
\ref{Sect:selfsimilar}).

If there is any important term in the energy equation other than
viscous dissipation and radiative losses, the discs are no longer
geometrically thin disks. The hydrostatic equilibrium
(\ref{eq:pressure}) reads
\[
\frac{H^2}{r^2}=\frac{\Omega^2}{\Omega^2+4\pi G \rho}\cdot \gamma\cdot \frac{c_s^2}{u_\phi^2}=\left(1-\eta_{\VSG}\right)\cdot \gamma\cdot\frac{c_s^2}{u_\phi^2}
\]
where $u_\phi=\Omega r$ is the azimuthal velocity, $\gamma$ the
adiabatic index, and $P=\gamma\rho c_s^2$ the speed of sound.
Disks are thin ($H/R\ll 1$) as long as they are cool (thermal
energy small compared to kinetic energy of the azimuthal flow).
Disks become hotter to reach the same $H/R$ ratio either for
$\gamma<\frac{5}{3}$ (ionisation or dissociation present) or for
$\eta_\VSG>0$ (see. $H/R(r)$ in Fig. \ref{fig:properties} for the
isothermal outer part of the $10^{-4}\Msun\yr^{-1}$ and
$10^{-3}\Msun \yr^{-1}$). The first effect is present in the
\citetalias{2004ApJ...603..383T} models whereas we set
$\gamma=\mbox{const}$. However \citetalias{2004ApJ...603..383T} do
not include vertical selfgravity which is the dominant vertical
force at these high accretion rates.

\label{lastpage}

\end{document}